\documentclass[MSNbibl,nameyear,seceqn,dvips]{arxstspdf}
\usepackage{tikz}
\usepackage{flushend}
\usepackage{stfloats}

\volume{29}
\issue{3}
\pubyear{2014}
\firstpage{367}
\lastpage{370}
\doi{10.1214/14-STS488} 
\referstodoi{10.1214/14-STS480}
\docsubty{FLA}

\makeatletter

\def\ci{\perp\!\!\!\perp}

\makeatother

\begin{document}
\begin{frontmatter}

\vspace*{12pt}\title{Causal Graphs: Addressing the~Confounding Problem Without Instruments~or~Ignorability}
\runtitle{Causal Graphs}

\begin{aug}
\author[a]{\fnms{Ilya} \snm{Shpitser}\corref{}\ead[label=e1]{i.shpitser@soton.ac.uk}}
\runauthor{I. Shpitser}

\affiliation{University of Southampton}

\address[a]{Ilya Shpitser is Lecturer in Statistics, Department of Mathematics, University of Southampton,
University Road, Building 54,
Southampton, Hampshire,
SO17 1BJ, United Kingdom \printead{e1}.}
\end{aug}


\end{frontmatter}

\section{Introduction}

I wish to congratulate Professor Imbens on a lucid and erudite review of
the instrumental variable literature.  The paper contrasts an econometric
view of instrumental variable models, where treatment confounding is due to
agents rationally choosing an optimal treatment for their situation, and
the statistical view, where treatment confounding arises due to noncompliance,
unobserved baseline differences between individuals, or other such issues.

While the paper does an admirable job describing the statistics view of
the instrumental variables based on the potential outcome model of Neyman
and Rubin, it does not much discuss the growing statistics literature on causal
graphical models, except to mention that causal graphs are a useful tool for
displaying the exclusion restriction assumption crucial for the use of
instrumental variables.

I would like to give a brief and hopefully complementary account of how
causal graphical models serve to clarify
and help address the issues of confounding (what Heckman calls the selection
problem) that make causal inference from
observational data such a challenging endeavor.

\section{Graphs as a General Method for Dealing with Confounding}

Causal inference in statistics has been greatly influenced by Neyman's idea of
explicitly representing interventions or forced treatment assignments on the
outcome (\cite{neyman23app}), and by Rubin's idea of using the stable unit
treatment value assumption (SUTVA) and ignorability assumptions to
equate potential outcome parameters with functionals of the observed data
(\cite{rubin74potential}).
Professor Imbens discusses these ideas at length in the paper.  The essence of
Rubin's method is that assumptions on potential outcome random variables allow
one to properly adjust for the presence of confounding.  Unfortunately, in
complex, possibly longitudinal settings it is not easy to see what assumptions
are needed, or whether it is even possible to identify parameters of interest
as functionals of observed data.  For this task, graphical causal models,
first used by Wright in the context of animal genetics
(\cite{wright21correlation}), and expanded into a general methodology for
causal inference by Spirtes, Glymour and Scheines (\citeyear{spirtes93causation}), Pearl
(\citeyear{pearl00causality}), Robins
(\citeyear{robins86new, robins97estimation}), and others
have proven to be invaluable.
%
\begin{figure*}
\begin{center}
\begin{tikzpicture}[>=stealth, node distance=1.2cm]
    \tikzstyle{format} = [draw, thick, circle, minimum size=5.0mm,
        inner sep=0pt]

        \begin{scope}
                \path[->, thick]
                        node[format] (a) {$A$}
                        node[format, above right of=a, gray] (c) {$C$}
                        node[format, below right of=c] (y) {$Y$}

                        (c) edge (a)
                        (c) edge (y)
                        (a) edge (y)

                        node[below of=c, yshift=-0.2cm] (l) {\footnotesize{(a)}}
                ;
        \end{scope}
        \begin{scope}[xshift=3.0cm]
                \path[->, thick]
                        node[format] (a) {$A$}
                        node[format, above right of=a] (c) {$C$}
                        node[format, below right of=c] (y) {$Y$}

                        (c) edge (a)
                        (c) edge (y)
                        (a) edge (y)

                        node[below of=c, yshift=-0.2cm] (l) {\footnotesize{(b)}}
                ;
        \end{scope}
        \begin{scope}[xshift=6.0cm]
                \path[->, thick]
                        node[format] (z) {$Z$}
                        node[format, right of=z] (a) {$A$}
                        node[format, above right of=a, gray] (c) {$C$}
                        node[format, below right of=c] (y) {$Y$}

                        (z) edge (a)
                        (c) edge (a)
                        (c) edge (y)
                        (a) edge (y)

                        node[below of=c, yshift=-0.2cm] (l) {\footnotesize{(c)}}
                ;
        \end{scope}
        \begin{scope}[xshift=10.2cm]
                \path[->, thick]
                        node[format] (a) {$A$}
                        node[format, above right of=a, gray] (c) {$C$}
                        node[format, below right of=c] (y) {$Y$}
                        node[format, right of=a, xshift=-0.35cm] (w) {$W$}

                        (c) edge (a)
                        (c) edge (y)
                        (a) edge (w)
                        (w) edge (y)

                        node[below of=c, yshift=-0.2cm] (l) {\footnotesize{(d)}}
                ;
        \end{scope}
\end{tikzpicture}
\end{center}
\caption{\textup{(a)} The standard problem of causal inference---an unobserved
confounder $C$, and possible approaches to the problem using observational
data.  \textup{(b)} Observing
the confounder and adjustment/stratification methods.
\textup{(c)} $Z$ as an instrumental variable.  \textup{(d)} A strong independent
mediator as an ``instrument'' for identification.}\label{fig:confounding}
\end{figure*}
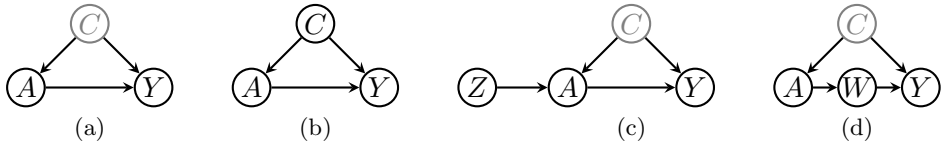

Consider Figure~\ref{fig:confounding}(a), where vertices represent random
variables of interest: a treatment $A$, an outcome $Y$, and a source of
unobserved confounding $C$
(lightly shaded in the graph to represented unobservability).
Following Neyman, we quantify the causal effect of $A$ on $Y$ by means of a
function of the distribution of the potential outcome $Y(a)$ ($Y$ after we
force $A$ to a value $a$).  For instance, we may use the average causal
effect (ACE): $E[Y(a)] - E[Y(a')]$, where $a$ is the active treatment value,
and $a'$ is the baseline treatment value.  We are interested in using
observed data to make inferences about such effects, which entails
dealing with confounding in some way.  The assumptions underlying this
graph which we will use can be expressed in terms of potential outcomes if
desired.  For example, the \emph{finest fully randomized causally interpretable
structured tree graph} (FFRCISTG) model of Robins (\citeyear{robins86new})
corresponding to this graph states that for all value assignments
$a$ and $c$ to $A$ and $C$, random variables $C$, $A(c)$ and $Y(a,c)$
are mutually independent, while the
\emph{nonparametric structural equation model with independent errors}
(NPSEM-IE) of Pearl (\citeyear{pearl00causality}) corresponding to this graph
states that for all value assignments $a,c$ and $c'$ to $A$ and $C$,
random variables $C$, $A(c)$ and $Y(c',a)$ are mutually independent.
Note that the former set of assumptions can be viewed
as a kind of mutual ignorability assumption derived from the graph, while
the latter set can be viewed as a mutual version of what Imai
called the \emph{sequential ignorability} assumption (\cite{imai10id}).
A general method for associating an arbitrary graph with sets of assumptions
on potential outcomes can be found, for instance, in the paper by \citet{thomas13swig}.
Thus, graphs are merely a visual representation of familiar potential outcome
models.  A~purely visual view, however, can prove quite helpful.

A common approach within the Rubin framework is to assume
that a conditional ignorability assumption $(Y(a) \ci A \mid C)$ holds.
Here, $\ci$ is the conditional independence symbol.
This assumption logically follows from the assumptions
defining the FFRCISTG model
of Figure~\ref{fig:confounding}(a).\footnote{The NPSEM-IE always makes at least
as many assumptions as the FFRCISTG model, and in many cases more.  Thus, any
assumption entailed by the FFRCISTG model of a graph is also entailed by the
NPSEM-IE for the same graph.}
Moreover, if $C$ is observed (represented graphically by
Figure~\ref{fig:confounding}(b), where $C$ is now normally shaded), this
assumption in turn entails that
the distribution of $Y(a)$ can be expressed as a functional of
the observed data via the adjustment formula
\[
p\bigl(Y(a)\bigr) = \sum_{c} p(Y \mid a,c) p(c)
\]
which in turn can be estimated by a variety of methods, including propensity
score methods (\cite{rosenbaum83propensity}),
inverse weighting methods (\cite{horvitz52weights}),
the parametric g-formula (\cite{robins87graphical})
or doubly robust methods (\cite{robins94estimation}).

If conditional ignorability is not a sensible assumption, or we cannot make use
of it due to strong sources of confounding that cannot be measured,
as is often the case in econometric applications, we may instead try to find
an instrument, which is shown graphically in Figure~\ref{fig:confounding}(c).
Here, the missing arrow from $Z$ to $Y$ represents the exclusion restriction
assumption namely that $Y(a,z) = Y(a,z')$ for any $a,z$ and $z'$,
and the lack of arrows from $C$ to $Z$ represents the assumption that
the instrument behaves as if randomly assigned: $Z \ci \{ A(z), Y(a,z) \}$
for any $a,z$.  These assumptions also logically follow from the assumptions
defining the FFRCISTG model for the graph in Figure~\ref{fig:confounding}(c).
As Professor Imbens discusses, given
these assumptions, one could obtain bounds on the effect, or using further
assumptions, obtain point identification.

It may be that we cannot make sure of the conditional ignorability assumption
(that is, large parts of $C$ are not observable),
and it is the case that a variable for which
the above instrumental assumptions hold cannot be found.
In this case, it is possible to
use the systematic representation of restrictions on potential outcomes given
by a graph to derive additional methods of attack on the confounding
problem which can complement the instrumental variable and conditional
ignorability approaches.

For example, it may be possible that a strong \emph{mediating} variable
for the effect of $A$ on $Y$ exists and is observable, and moreover,
the mechanism by which this mediation happens is independent of the source
of the confounding between the treatment $A$ and outcome $Y$, given the
treatment $A$.  This situation is shown in Figure~\ref{fig:confounding}(d).
In terms of assumptions on potential outcomes, such a strong mediator between
$A$ and $Y$ is represented as stating that $Y(w,a) = Y(w,a')$ for all $w,a,a'$.
In words, $A$ has no direct effect on $Y$ once we fix $W$ to any value $w$.
The unconfoundedness of the mediator may be represented as stating that
$\{ Y(w), A \} \ci W(a)$.  It is easy
to see that the former assumption is a kind of exclusion restriction
corresponding to the absence of a directed arc from $A$ to $Y$, and the latter
is a kind of ignorability assumption, which corresponds to the absence of
an arc from $C$ to $W$, and the absence of other sources of confounding
between $W$ and $A$ and $Y$.  Professor Imbens discusses these kinds of assumptions
in the context of instrumental variables in Sections~5.1 and 5.2.
Pearl has shown a result that is equivalent to stating that given a version of
SUTVA, and above assumptions, the following \emph{front-door formula} holds:
\begin{eqnarray}\label{eqn:front-door}
\hspace*{24pt}p\bigl(Y(a)\bigr) = \sum_{w} p(w \mid a) \sum
_{a'} p\bigl(Y \mid w, a'\bigr) p
\bigl(a'\bigr).
\end{eqnarray}
As before, the above assumptions logically follow from the assumptions defining
the FFRCISTG model for Figure~\ref{fig:confounding}(d).

A (slightly contrived) example of a situation represented in Figure~\ref{fig:confounding}(d) is as follows.  We may suspect that consumption of
skyr (a kind of Icelandic dairy product) is protective against stomach cancer by
means of a particular type of flora found in skyr.  However, we suspect
those with Icelandic citizenship may both consume more skyr than average, and
have different rates of stomach cancer than the general population.  If we
cannot observe citizenship status, but we can do a simple test for the presence
of the protective flora, and moreover, we suspect the protective causal
mechanism is not influenced by the confounding variable directly but only
through skyr consumption, and moreover, this mechanism mediates
\emph{all} of the effect of skyr consumption, then we can find a way to express
the causal effect of skyr consumption on incidence of stomach cancer using
observational data via (\ref{eqn:front-door}).

Since graphs are merely a systematic \emph{visual} way of arranging information
on potential outcomes, they have proven extremely helpful for generating
solutions to problems posed by confounding in very general settings.
For instance, \citet{tyler13on}
used graphs to show that many informal
definitions for what a ``confounder'' is in the literature are ``incorrect''
in the sense of not agreeing with intuition in given examples, and not obeying
certain natural properties we expect a confounder to obey, while a definition
that is ``correct'' in this sense is fairly subtle.

Furthermore, a mediator-based approach resulting in (\ref{eqn:front-door})
has been generalized using causal graphs to a fully general method of
``deconfounding'' which can successfully be applied in complex longitudinal
settings even when no standard ignorability assumptions can be used,
and no good instruments can be found.

\begin{figure}
\begin{center}
\begin{tikzpicture}[>=stealth, node distance=1.2cm]
    \tikzstyle{format} = [draw, thick, circle, minimum size=5.0mm,
        inner sep=0pt]

        \begin{scope}
                \path[->, thick]
                        node[format] (a) {$A$}
                        node[format, right of=a] (b) {$B$}
                        node[format, right of=b] (c) {$C$}
                        node[format, right of=c] (d) {$D$}
                        node[format, right of=d] (e) {$Y$}

                        (a) edge (b)
                        (b) edge (c)
                        (b) edge[bend right] (d)
                        (c) edge (d)
                        (d) edge (e)
                        (c) edge[bend right] (e)
                        (b) edge[bend left=45] (e)

                        (a) edge[<->, thick, bend left] (c)
                        (c) edge[<->, thick, bend left] (e)
                        (a) edge[<->, thick, bend right=45] (e)
                        (a) edge[<->, thick, bend left=45] (d)
                ;
        \end{scope}
\end{tikzpicture}
\end{center}
\caption{A longitudinal study with exposures $B,D$, outcome $Y$ and multiple
sources of unobserved confounding.}
\label{fig:jamie}
\end{figure}
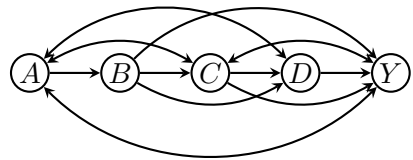

Consider a hypothetical longitudinal study represented by the causal graph
shown in Figure~\ref{fig:jamie}, where bidirected arrows represent the presence
of some hidden common cause.  For instance, a bidirected arrow from $A$ to $C$
means there is a hidden common cause of $A$ and $C$.  In this study,
$B$ and $D$ are administered treatments, $Y$ is the outcome, $A$ is
an observed baseline confounder, and $C$ is an intermediate health measure.
These variables are confounded, but in a very particular way displayed
by bidirected arrows in the graph.  For instance, there
is no hidden common cause of $B$ and any other variable, and $D$ has
a hidden cause in common only with~$A$.
%

We are interested in the total effect of the drug on the outcome, which can
be obtained from the distribution of $Y(b,d)$, in order to better determine the
optimal treatment assignment.  Robins (\citeyear{robins86new})
gave a general method for expressing $Y(b,d)$ as a functional
of the observed data given that an assumption called sequential ignorability
(different from one in \cite{imai10id}) holds.  Sequential ignorability on the observable variables happens
not to be implied by either the FFRCISTG model, or the NPSEM-IE of any underlying hidden variable DAG consistent with
Figure~\ref{fig:jamie}, but it can be shown that assumptions implied by these
models can be used to derive the following identity:
\begin{eqnarray}\label{eqn:complex}
p\bigl(Y(b,d) = y\bigr) &=& \sum_c \biggl( \sum
_{a} p(y,d,c \mid b,a) p(a) \biggr)\nonumber\\[-8pt]\\[-8pt]
&&{} \cdot
\frac{\sum_{a} p(c \mid b,a) p(a) }{
 \sum_{a} p(d,c \mid b,a) p(a) }. \nonumber
\end{eqnarray}

Just as in the previous cases, it is possible to derive in a systematic
way a list of restrictions on potential outcomes over observable variables
implied by the graph in Figure~\ref{fig:jamie}, and use this list to
derive (\ref{eqn:complex}) without referring to the graph at all.
However, without the help of the graph
it is not so easy to see what this list of restrictions might look like, or how
the derivation based on this list might proceed.  In more complex longitudinal
settings, with many more than 5 variables the problem becomes even more severe.

A general method for identifying potential outcome distributions as functionals
of observed data for causal graphs was given in the paper by \citet{tian02on}, and was proven
complete (in the sense of only failing on nonidentifiable distributions)
by \citeauthor{shpitser06id} (\citeyear{shpitser06idc}, \citeyear{shpitser06id}, \citeyear{shpitser07hierarchy}), \citet{huang06do}.
While it is possible to rederive these results
in terms of assumptions on potential outcomes, graph theory provided
mathematical terminology and intuitions that proved crucial in practice
for deriving these general ``deconfounding'' results.


\section{Conclusion}

Professor Imbens' excellent review shows that in the context
of economics, confounding arises due to the agent's decision algorithm, while
in the context of statistics confounding arises for other reasons (for
instance, due to lack of compatibility among patients in an observational
study).  Instrumental variables are an important technique for dealing with
confounding in causal inference, though alternative methods involving
stratification were developed under the potential outcome model of Rubin and
Neyman.


More general lines of attack on the problem of confounding
were developed using graphical causal models, first used by Wright in
genetics, and extended into a general model by Spirtes, Glymour and Scheines
(\citeyear{spirtes93causation}), Pearl (\citeyear{pearl00causality}), Robins
(\citeyear{robins86new}, \citeyear{robins97estimation})
and others.  These methods allow ``deconfounding'' of the problem even in
cases where confounders cannot be observed, and good instrumental variables
do not exist.




\end{document}